\documentclass[journal,a4paper]{IEEEtran}

\usepackage{cite}
\usepackage[english]{babel}
\usepackage[utf8]{inputenc}
\usepackage[cmex10]{amsmath}
\usepackage{graphicx}
\usepackage{xcolor}
\usepackage{siunitx}
\usepackage[caption=false,font=footnotesize]{subfig}

\definecolor{MyRed}{rgb}{1.0,0.0,0.0}

\begin{document}

\title{Deep Learning-based Initialization of \\
Iterative Reconstruction for Breast Tomosynthesis}

\author{Koen~Michielsen, Nikita~Moriakov, Jonas~Teuwen, and~Ioannis~Sechopoulos%
\thanks{K. Michielsen, N. Moriakov, J. Teuwen, and I. Sechopoulos are with the dept. of Radiology, Nuclear Medicine, and Anatomy of Radboudumc, Nijmegen, The Netherlands.}%
\thanks{J. Teuwen is with the Department of Radiation Oncology, Netherlands Cancer Institute, Amsterdam, The Netherlands}%
\thanks{I. Sechopoulos is with the Dutch Expert Center for Screening (LRCB), Nijmegen, The Netherlands.}%
}

\maketitle
\thispagestyle{empty}
\pagestyle{empty}

\begin{abstract}
Reconstruction of digital breast tomosynthesis is a challenging problem due to the limited angle data available in such systems.
Due to memory limitations, deep learning-based methods can help improve these reconstructions, but can not (yet) attain sufficiently high resolution. In addition to this practical issue, questions remain on the possibility of such models introducing 'ghost' information from the training data that is not compatible with the projection data.
To take advantage of some of the benefits of deep learning-based reconstructions while avoiding these limitations, we propose to use the low resolution deep learning-based reconstruction as an initialization of a regular high resolution iterative method.

The network was trained using digital phantoms, some based on a mathematical model and some derived from patient dedicated breast CT scans. The output of this network was then used as initialization for 10\,000 iterations of MLTR for nine patient based phantoms that were not included in the training. The same nine cases were also reconstructed without any initialization for comparison.

The reconstructions including initialization were found to reach a lower mean squared error than those without, and visual inspection found much improved retrieval of the breast outline and depiction of the skin, confirming that adding the deep learning-based initialization adds valuable information to the reconstruction.
\end{abstract}


\section{Introduction}
Digital breast tomosynthesis (DBT) is a limited angle tomography modality dedicated to breast imaging. It improves on digital mammography by providing a pseudo-3D reconstruction with high in-plane resolution, but low resolution in the third dimension.
This lower resolution is caused by the incomplete angular sampling, resulting in missing data in the Fourier domain. This also means that the reconstruction problem is under-determined, and that the result of an iterative reconstruction will depend on its starting point. Therefore, a good initialization not only helps increase convergence speed, but, in DBT, also improves the reconstruction itself \cite{Michielsen2015-uh}.

Recently, we developed a deep learning-based reconstruction for a simplified 2D version of the DBT limited angle problem\cite{Moriakov2019-qm}. With it, we were able to reconstruct the distribution of the glandular tissue in the breast to an accuracy that allowed for the estimation of the individual patient dose to within $20\%$ and of the total glandular tissue to within $3\%$.
Extending this work to 3D is a significant challenge due to the difficulty of fitting a complex convolutional neural network adapted to the large data sizes in DBT within GPU memory. Due to these constraints, for now it is only possible to perform these reconstructions at low in-plane resolutions.

However, given the promising results, in terms of increased vertical resolution and decreased limited angle artifacts, obtained with our deep learning-based reconstruction, here we examine if it could be introduced as a starting point for an iterative reconstruction method. If the two reconstruction methods are compatible, the iterative algorithm would not change the low resolution information present in the initialization, and only fill in the high resolution details.

\section{Materials \& Methods}
The experimental setup and patient-based phantoms used to train the deep learning network and evaluate the effect of the initialization are described in section~\ref{sec:phantom}, and the metrics used for the evaluation in section~\ref{sec:eval}. The deep learning reconstruction is presented in section~\ref{sec:dl_recon} and the iterative reconstruction in section~\ref{sec:mltr}.

\subsection{Phantom \& Acquisition}
\label{sec:phantom}
\begin{figure}
\centering
	{\includegraphics[width=0.24\textwidth]{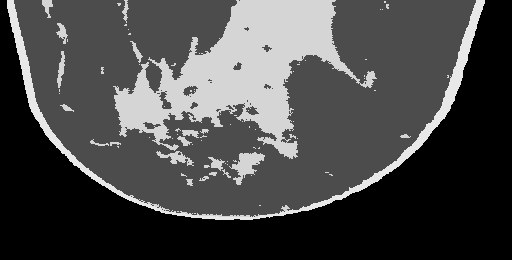}}%
    {\includegraphics[width=0.24\textwidth]{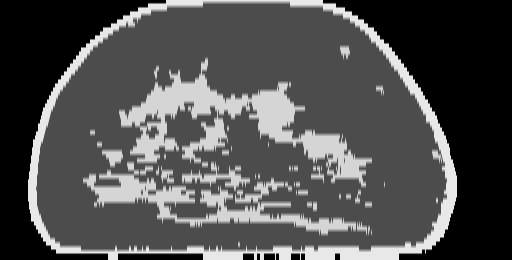}}%
    \caption{Axial and coronal views of a patient-based phantom. It contains three tissue types: skin at the edge, fibro-glandular structures near the center, and the remainder is adipose tissue.}%
    \label{fig:phantom}
\end{figure}

We mixed two sets of breast phantoms to both train the deep learning reconstruction and evaluate the follow-up iterative reconstruction. The first set contained 52 statistically defined anthropomorphic phantoms\cite{Lau2012-uc}, while the second set consisted of 90 phantoms based on patient scans obtained using a dedicated breast CT system.   
To generate the phantoms in this second set, the scans were segmented into skin, adipose tissue, and fibro-glandular tissue\cite{Caballo2018-eg}, and then digitally compressed to obtain a phantom that could be used for DBT imaging\cite{Fedon2018-ds}. An example of the end result is shown in figure~\ref{fig:phantom}. We randomly chose 110 samples across the two sets for training and validating the model, with the remaining phantoms reserved for testing. For ease of use, both data sets were converted to an isotropic voxel spacing of \SI{0.25}{\milli\meter}

X-ray projection images were simulated in a geometry based on the Mammomat Inspiration system (Siemens Healthcare, Forchheim, Germany) \cite{Sechopoulos2013-rv}, with a few minor modifications: the detector pixel spacing was set to \SI{0.250}{\milli\meter}, and the detector cover and compression paddle were not included.

A simplified physical model, shown in equation~(\ref{eq:forward_model}), was used in the simulation:
\begin{equation}
	\hat{y}_i \left( \vec{\mu} \right) = b_i e^{-\sum_j l_{ij} \mu_j} \mathrm{,}
	\label{eq:forward_model}
\end{equation}
with $\vec{\mu}$ being the phantom linear attenuation coefficients, $l_{ij}$ the intersection length between ray~$i$ and voxel~$j$, $b_i$ the unattenuated photon count at pixel~$i$, and $\hat{y}$ the resulting projection data.
Linear attenuation coefficients for the phantom were calculated for a mono-energetic x-ray beam at \SI{20}{\kilo\electronvolt} with the software of Boone and Chavez\cite{Boone1996-tu} using the tissue compositions published by Hammerstein et al.\cite{Hammerstein1979-uc}.
Blank scan value $\vec{b}$ was set to 16\,000 photons to obtain the noise equivalent to a relatively high dose scan.

\subsection{Deep Learning Reconstruction}
\label{sec:dl_recon}
The algorithm used to compute the low resolution 3D DBT reconstructions, referred to as DBToR\nobreakdash-X for now, is a modified, memory-optimized version of the DBToR algorithm that was originally developed for 2D slice-wise DBT reconstruction\cite{Moriakov2019-qm}, and was adapted for the purpose of 3D reconstruction with the realistic projection geometry described in section~\ref{sec:phantom}.

The DBToR algorithm takes the raw sinogram and the measured compressed breast thickness as inputs and performs the reconstruction by a neural network, consisting of primal and dual reconstruction blocks\cite{Adler2018-hc}. The primal and dual reconstruction blocks are given by small Convolutional Neural Networks (CNNs) that operate in primal (reconstruction) and dual (projection) spaces, respectively. Primal and dual blocks are connected to each other by linear projection and backprojection operators, which implement the corresponding projection geometry. 
This results in an `unrolled iterative scheme', where the neural network starts with zero as the initial guess and iteratively improves the reconstruction by applying CNNs in primal and dual spaces. Since all operations are differentiable, the network can be trained end-to-end to minimize the $L^2$ reconstruction loss on the training set. The inclusion of the compressed breast thickness information in the algorithm significantly improves the reconstruction quality. 
The DBToR\nobreakdash-X version of the algorithm was implemented in PyTorch, and uses the Operator Discretization Library (ODL) with a custom projection geometry to match the DBT acquisition setup. The model was trained to minimize $L^2$ loss of the reconstruction on the training set, but despite the included memory optimizations, it was not possible to perform this training at full resolution, and output was restricted to a voxel size of \SI{0.75}{\milli\meter}$\times$\SI{0.75}{\milli\meter}$\times$\SI{1.0}{\milli\meter}.

Nine of the patient-based phantoms in the test set were randomly selected and the matching DBToR\nobreakdash-X reconstructions were calculated. These were first up-sampled to full resolution without interpolation and then used as initial guess for the MLTR reconstruction described in the next section.

\subsection{Maximum Likelihood Reconstruction}
\label{sec:mltr}
To evaluate the usefulness of the low resolution deep learning reconstruction as initialization of a high resolution iterative method we continued the reconstruction with the MLTR algorithm\cite{Nuyts1998-ea}.
This method maximizes the log-likelihood cost function $\mathcal{L}$ in equation~(\ref{eq:likelihood}), 
with $y_i$ the measured sinogram and $\hat{y}_i$ the forward model from equation~(\ref{eq:forward_model}). This is done by iteratively updating reconstruction volume $\vec{\mu}^{(n)}$.

\begin{equation}
	\mathcal{L} (\vec{\mu}) = \sum_i y_i \ln \hat{y}_i - \hat{y}_i 
    \label{eq:likelihood}
\end{equation}

The update step at each iteration is calculated by taking the second order Taylor expansion of $\mathcal{L}$ at that point and further approximating this result by a separable quadratic surrogate function for which the optimum can then be reached in a single update, which is shown in equation~(\ref{eq:mltr}).

\begin{equation}
	\mu_j^{(n+1)} = \mu_j^{(n)} + \frac{\sum_i l_{ij} \left( \hat{y}_i^{(n)} - y_i \right)}
	{\sum_i l_{ij} \hat{y}_i^{(n)} \sum_k l_{ik}} 
	\label{eq:mltr}
\end{equation}

\subsection{Evaluation}
\label{sec:eval}
Nine of the patient-based phantoms were selected for reconstruction with 10\,000 iterations, once without any initialization (all zero values) and once starting from the deep learning-based reconstruction, which was first up-sampled to the full resolution (\SI{0.25}{\milli\meter}$\times$\SI{0.25}{\milli\meter}). Except for the initialization, there was no difference between the reconstructions.
Intermediary results were saved after iterations 1 through 10, every $10^{\mathrm{th}}$ iteration until iteration 100, every $100^{\mathrm{th}}$ iteration until iteration 1\,000, and then every $500^{\mathrm{th}}$ iteration.

To evaluate the reconstructions, we calculated the log-likelihood cost function and the mean squared error (MSE) for all intermediary results and visually inspected a small number of cases.

\section{Results}
\label{sec:results}
\begin{figure*}
\centering
	{\includegraphics[width=0.25\textwidth]{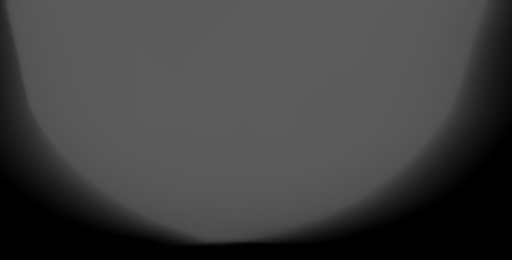}}%
    {\includegraphics[width=0.25\textwidth]{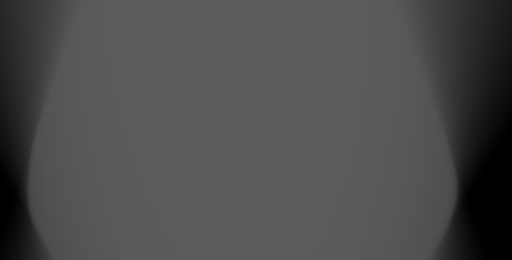}}%
	{\includegraphics[width=0.25\textwidth]{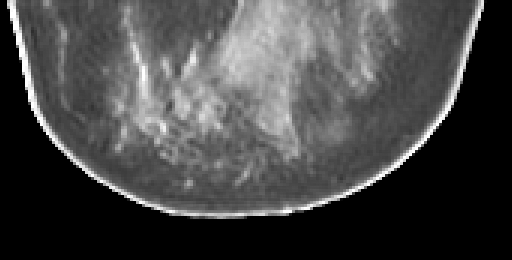}}%
	{\includegraphics[width=0.25\textwidth]{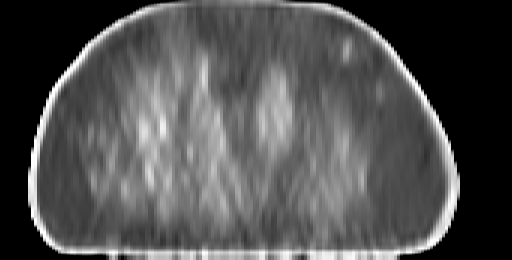}}\\
	
	{\includegraphics[width=0.25\textwidth]{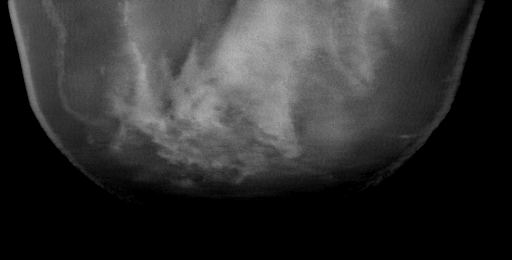}}%
	{\includegraphics[width=0.25\textwidth]{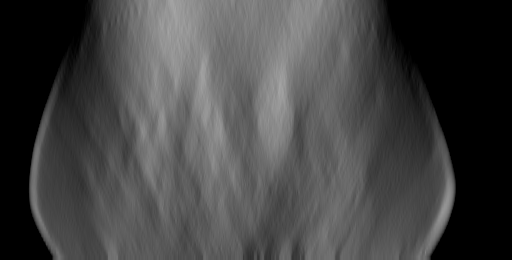}}%
	{\includegraphics[width=0.25\textwidth]{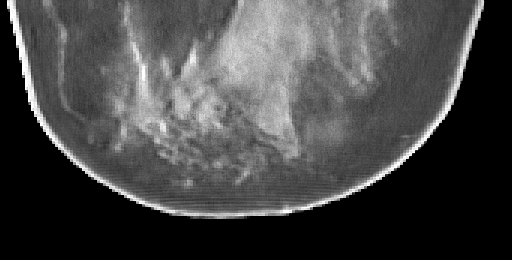}}%
	{\includegraphics[width=0.25\textwidth]{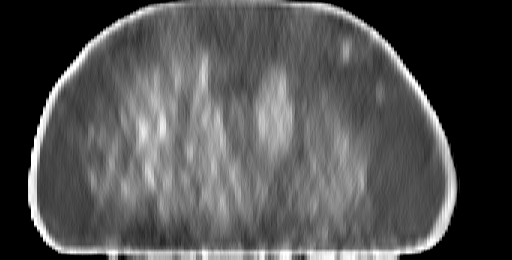}}\\
	
	{\includegraphics[width=0.25\textwidth]{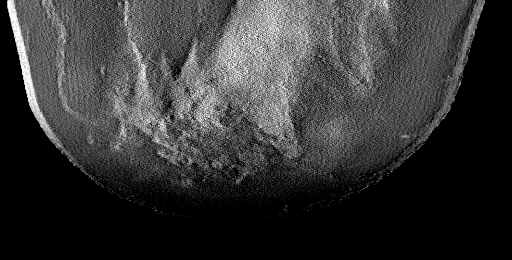}}%
	{\includegraphics[width=0.25\textwidth]{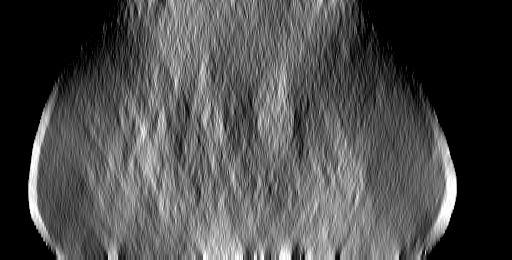}}%
	{\includegraphics[width=0.25\textwidth]{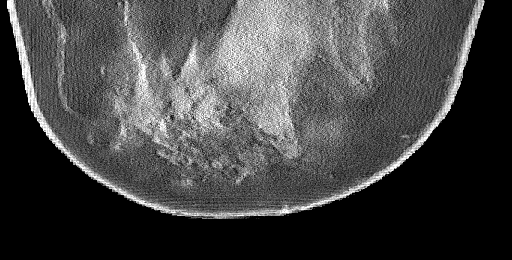}}%
	{\includegraphics[width=0.25\textwidth]{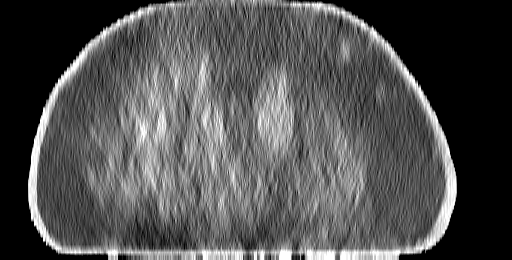}}\\
	
	{\includegraphics[width=0.25\textwidth]{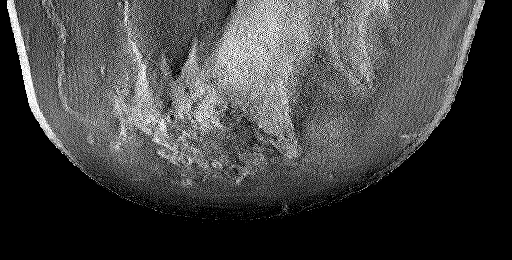}}%
	{\includegraphics[width=0.25\textwidth]{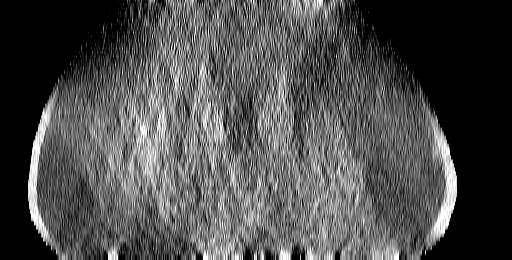}}%
	{\includegraphics[width=0.25\textwidth]{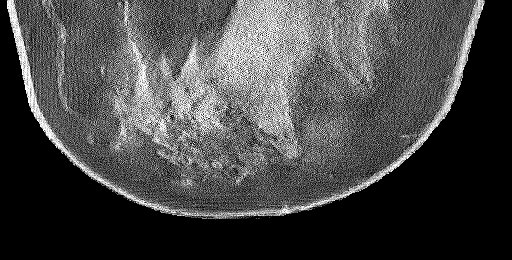}}%
	{\includegraphics[width=0.25\textwidth]{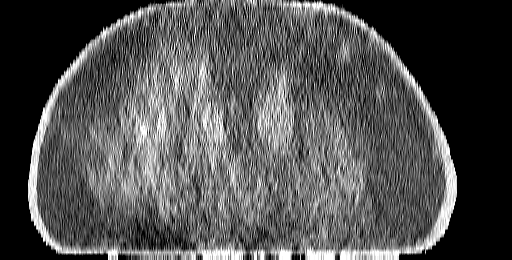}}\\
	
    {\includegraphics[width=0.25\textwidth]{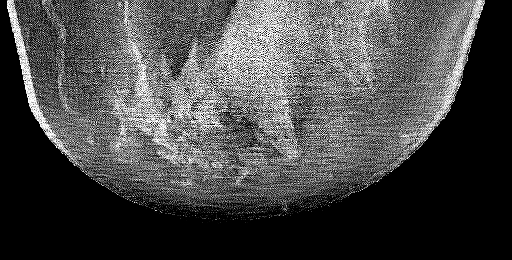}}%
    {\includegraphics[width=0.25\textwidth]{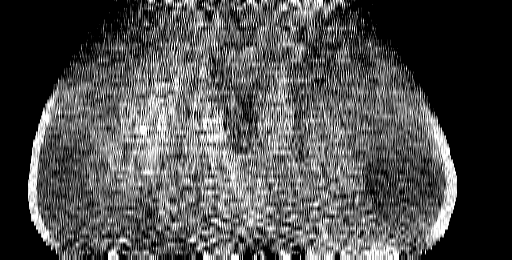}}%
    {\includegraphics[width=0.25\textwidth]{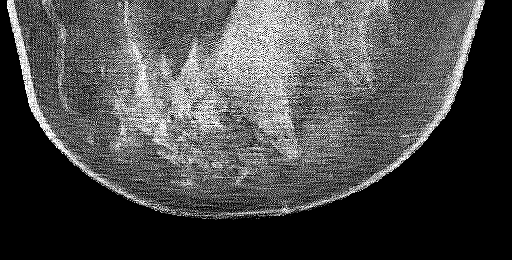}}%
    {\includegraphics[width=0.25\textwidth]{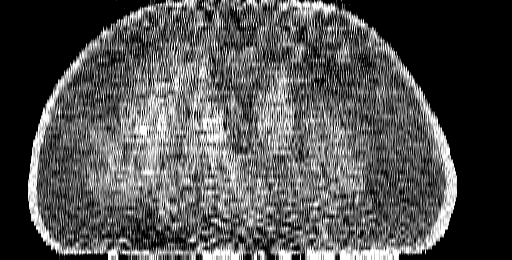}}
    \caption{Axial and coronal reconstruction views after iterations 1, 10, 100, 1\,000, and 10\,000 (top to bottom). Reconstructions without initialization are shown on the left hand side, and reconstructions with the deep learning-based reconstruction on the right hand side. The window is set between \SI{0.04}{\per\milli\meter} and \SI{0.09}{\per\milli\meter}, except for the first iteration without initialization (top left) where the window is set between \SI{0.00}{\per\milli\meter} and \SI{0.04}{\per\milli\meter}.}%
    \label{fig:results}
\end{figure*}

Figure~\ref{fig:results} shows axial and coronal views of reconstructions with and without the deep learning-based initialization after 1, 10, 100, 1\,000, and 10\,000 iterations. These reconstructions are based on the phantom shown in figure~\ref{fig:phantom}.
In the reconstruction without initialization, attenuation in the anterior and superior regions remains underestimated and the skin is not visualized even after 10\,000 iterations, while the skin is clearly depicted from the start, and remains clearly visible, in the deep learning-initialized reconstruction.

\begin{figure}
\centering
	{\includegraphics[width=0.48\textwidth]{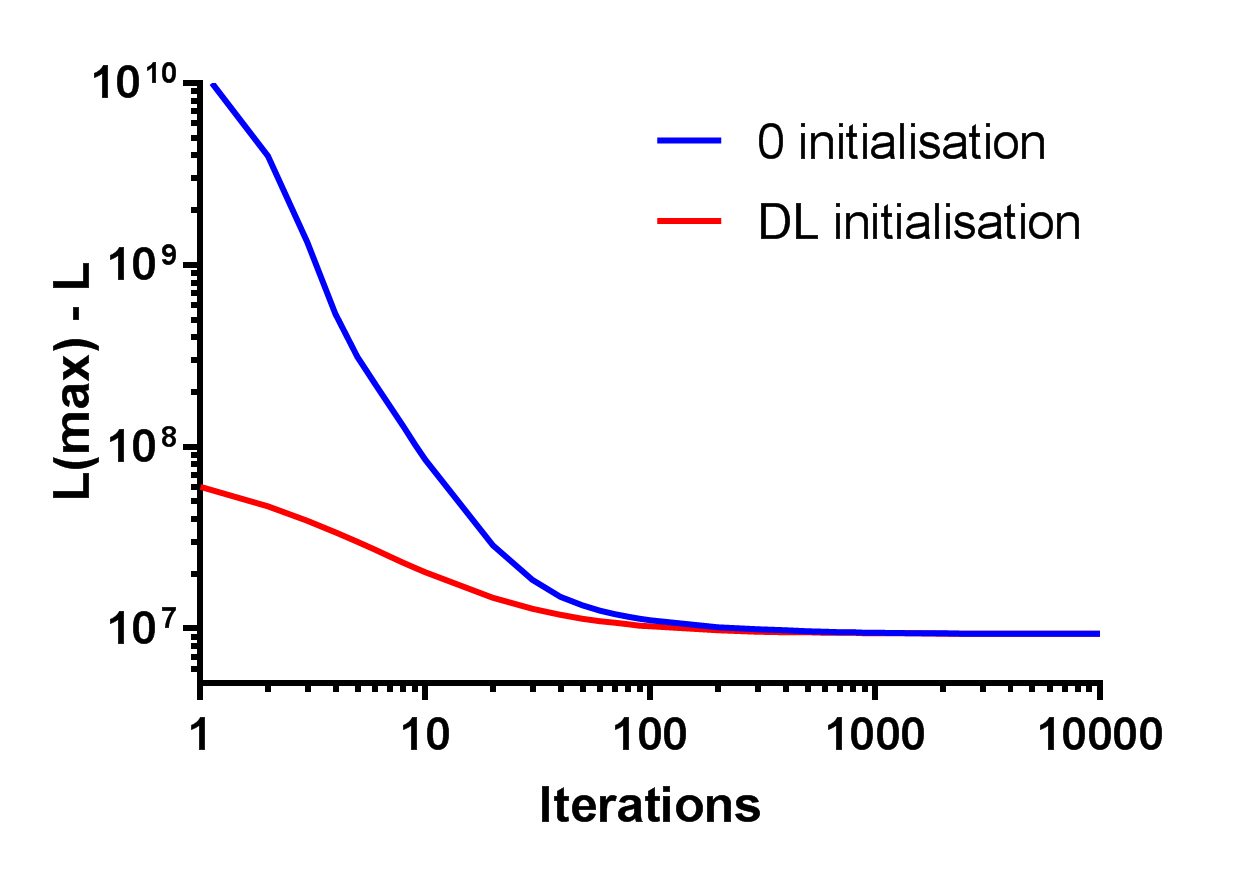}}%
    \caption{Difference between the maximum log-likelihood and the log-likelihood cost as a function of iterations.}%
    \label{fig:051_llik}
\end{figure}

\begin{figure}
\centering
	{\includegraphics[width=0.48\textwidth]{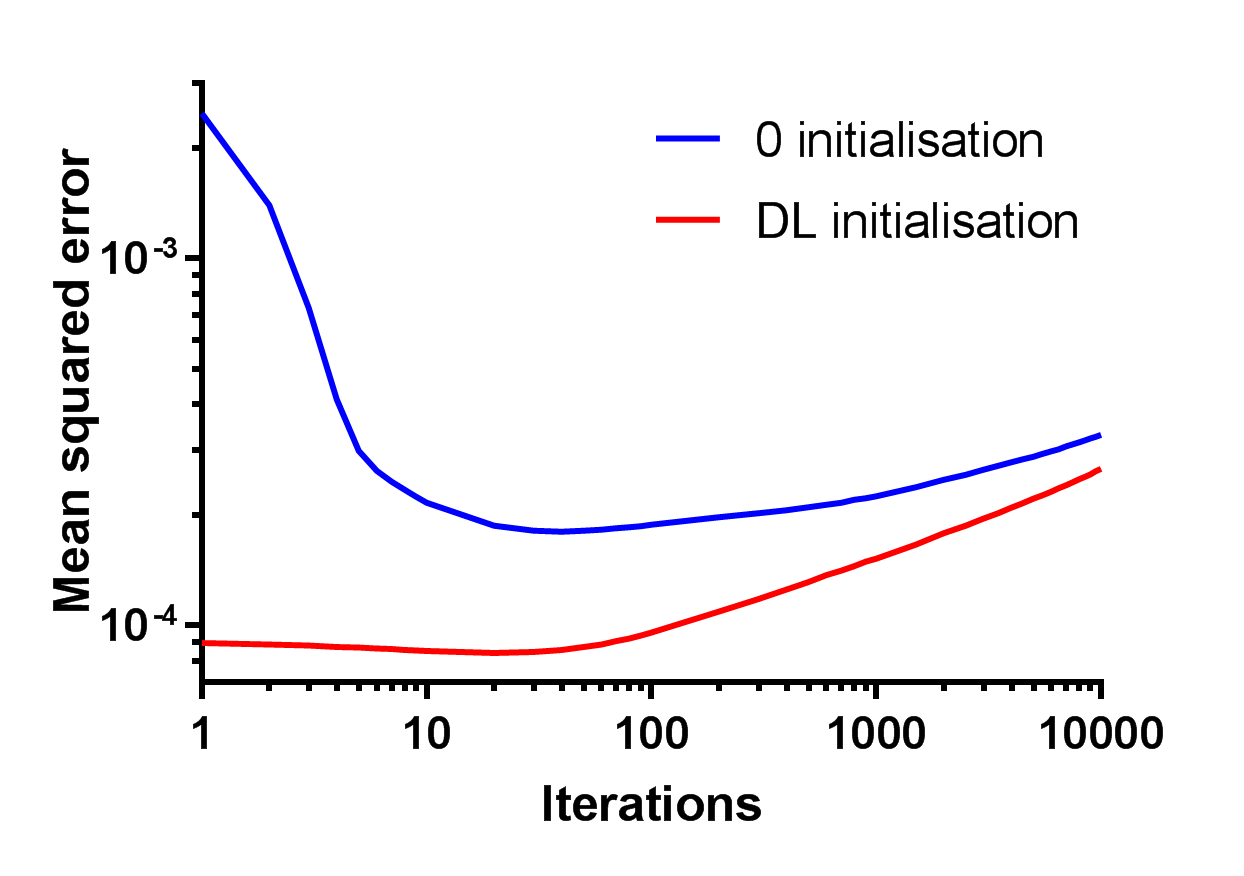}}%
    \caption{Mean squared error of the reconstructions as a function of iterations.}%
    \label{fig:051_ssqd}
\end{figure}

Figures~\ref{fig:051_llik} and~\ref{fig:051_ssqd} plot the evolution of the log-likelihood cost function and MSE, respectively, for the same phantom, which are representative of the behavior seen for these two metrics with all 9 tested phantoms.
The mean squared error for all 9 phantoms and both reconstructions reaches a minimum between iterations 10 and 100 and then starts increasing again. Meanwhile, the minimum mean squared error of the reconstructions with initialization is always lower than that of the reconstructions without initialization. 
After 50 iterations, the average MSE was \SI{177d-6}{\per\milli\meter\squared} for the reconstructions without initialization and \SI{101d-6}{\per\milli\meter\squared} for those with initialization.

Log-likelihood was found to be monotonously increasing for all cases, but was not found to be consistently higher for reconstructions with or without the initialization. Specifically, in four cases it was higher for the reconstructions with initialization while for five cases it was higher for the reconstructions without initialization.

\begin{figure}
\centering
	{\includegraphics[width=0.24\textwidth]{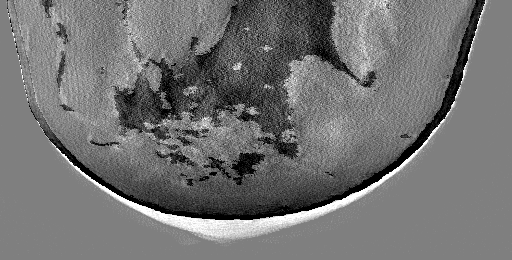}}%
    {\includegraphics[width=0.24\textwidth]{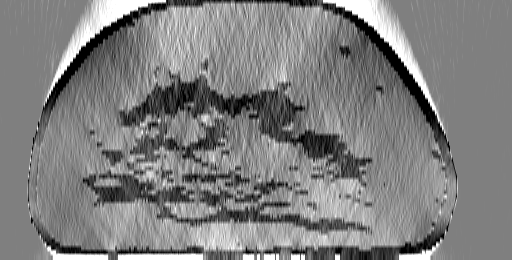}}\\%
    {\includegraphics[width=0.24\textwidth]{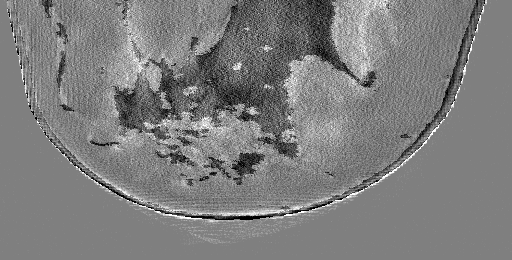}}%
    {\includegraphics[width=0.24\textwidth]{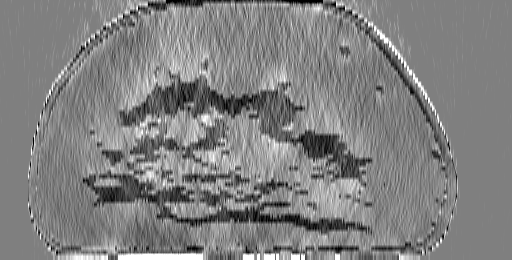}}\\%
	{\includegraphics[width=0.24\textwidth]{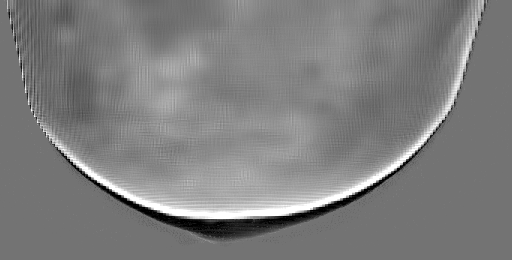}}%
    {\includegraphics[width=0.24\textwidth]{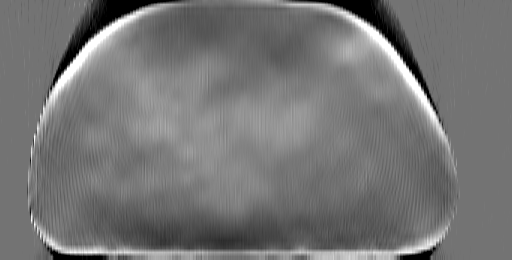}}%
    \caption{Difference between the reconstruction without initialization and the ground truth (top row), difference between the reconstruction with initialization and the ground truth (middle row), and difference between both reconstructions (bottom row), all after 100 iterations. The window is set between \SI{-0.03}{\per\milli\meter} and \SI{0.03}{\per\milli\meter}.}%
    \label{fig:delta}
\end{figure}

Figure~\ref{fig:delta} shows the difference between both reconstructions and the difference with the ground truth after 100 iterations. 
Compared to the ground truth, the lack of resolution due to the limited angle acquisition is clearly visible in the distribution of fibro-glandular tissue in the coronal slices, for both methods. However, the skin line is well depicted in the initialized reconstruction, with only a minimal outward shift relative to the ground truth.
The comparison between the two methods highlights the differences in visualization of the skin and also hints at a more accurate distribution of the fibro-glandular tissue in the initialized reconstruction, with the subtraction indicating higher attenuation in the center and lower attenuation in the peripheral regions compared to the reconstruction without initialization.

\section{Discussion \& Conclusion}
Our results clearly show the positive impact gained by including a low resolution deep learning-based reconstruction as initialization of an iterative reconstruction algorithm. The information the initialization adds is compatible with the data, since it was not removed even after 10\,000 iterations, but could also not be retrieved from the data by the iterative reconstruction.
It is not clear from the metrics in figures~\ref{fig:051_llik} and~\ref{fig:051_ssqd} whether or not both reconstructions are converging to the same result, possibly because both methods are starting to over-fit to the noise at that point.
At least the visual comparison in figure~\ref{fig:results} indicates that the visualization of the skin will not appear in the un-initialized reconstruction, and will not disappear from the initialized reconstruction, and thus both seem to converge to a different solution.
For all practical purposes, reconstructions will be stopped well before reaching 10\,000 iterations, and the point of convergence is less important than which solution can be found in a relatively short time. With that in mind, the deep learning based reconstruction substantially improves the result.

At this point in time the main reason to combine DBToR\nobreakdash-X with MLTR is that the deep learning based method cannot reach the high resolution needed for breast imaging, due to memory constraints in the training process, and thus the regular iterative method is needed to fill in the high resolution details.
However, we believe this approach remains valid even if DBToR\nobreakdash-X could process reconstructions at full resolution. Because it behaves as a reconstruction combined with some anatomical prior knowledge, it could place structures that are not compatible with the data, especially since it is not clear how much weight is put on the prior knowledge relative to the data fidelity. Combination with a regular iterative method would ensure that such incompatible structures would be removed from the image.

Some modifications would be needed to extend the method to make it applicable to patient data. DBToR-X would need pre-corrected projection data as input, but, this should not be problematic, since filtered backprojection has the same requirement. The main remaining question is, then, how relevant the implicit anatomical prior information learned from the (patient-based) digital phantoms would be, and if it would still help reconstruction quality.

\vspace{3mm}
To conclude, we found that the reconstructions that included initialization reached a lower mean squared error than those without, and visual inspection found considerably improved retrieval of the breast outline and depiction of the skin, confirming that adding the deep learning-based initialization adds valuable information to the reconstruction.

\vspace{3mm}
{\it This work was presented at the 6th International \mbox{Conference} on Image Formation in X-Ray Computed Tomography ({CT\nobreakdash-Meeting}), August 6, 2020, Regensburg, Germany.}



\bibliographystyle{IEEEtran}
\bibliography{IEEEabrv,paperpile.bib}

\end{document}